\documentclass[twocolumn,tighten,dvipsnames]{aastex631}
\usepackage{amsmath}
\usepackage{xspace}
\usepackage{booktabs}
\usepackage{bm}
\usepackage{hyperref}

\newcommand{\SPO}{SP\"O\xspace}
\newcommand{\ipole}{\texttt{ipole}\xspace}
\newcommand{\SGRA}{Sgr~A*\xspace}

\newcommand{\Rh}{R_\mathrm{high}}
\newcommand{\Rl}{R_\mathrm{low}}
\newcommand{\betac}{\beta_\mathrm{crit}}
\newcommand{\spin}{a_\mathrm{spin}}

\newcommand{\PSPA}{P_\mathrm{S}/P_\mathrm{A}}

\defcitealias{2019ApJ...875L...1E}{EHTC et~al. 2019a}
\defcitealias{2019ApJ...875L...5E}{EHTC et~al. 2019b}
\defcitealias{2022ApJ...930L..12E}{EHTC et~al. 2022a}
\defcitealias{2022ApJ...930L..16E}{EHTC et~al. 2022b}

\begin{document}

\title{General Relativistic Radiative Transfer with a Physics-Motivated Electron Temperature Model}

\author{Yi Tung Tsang}
\affiliation{Department of Physics, The Chinese University of Hong Kong, Sha Tin, N.T., Hong Kong}

\author[0000-0003-0433-3585]{Kaushik Satapathy}
\affiliation{Instituto de Plasmas e Fus\~{a}o Nuclear, Instituto Superior Técnico, Av. Rovisco Pais 1, 1049-001 Lisboa}



\author[0000-0001-6337-6126]{Chi-kwan Chan}
\affiliation{Steward Observatory and Department of Astronomy, University of Arizona, 933 N.~Cherry Ave., Tucson, AZ~85721, USA}
\affiliation{Office of Responsible AI, University of Arizona, 750 N.~Cherry Ave, Tucson, AZ~85719, USA}
\affiliation{Program in Applied Mathematics, University of Arizona, 617 N.~Santa Rita, Tucson, AZ~85721, USA}

\begin{abstract}
  Accurate modeling of electron thermodynamics in black hole accretion
  disks remains a long-standing challenge.
  The commonly used parametric $\Rl$-$\Rh$ model does not explicitly
  model the microphysics of collisionless plasma in accretion flows.
  Without physical constraints on its parameters, large simulation
  surveys are required to infer black hole properties.
  Here, we explore the observational consequences of a physically
  motivated model from \citet*{2024ApJ...969..100S}, hereafter \SPO,
  implemented in the ray-tracing code \ipole.
  We compare synthetic images produced by this model to those from a
  best-fit SANE $\Rl$-$\Rh$ model and assess their consistency with
  EHT observations of \SGRA.
  We find that the choice of electron model significantly affects the
  optical depth of the accretion flow.
  The \SPO model concentrates emission closer to the black hole and
  produces a more asymmetric, optically thinner accretion disk.
  These changes affect both polarization and variability.
  Our results reaffirm that the SANE model alone is insufficient to
  explain the EHT observations of \SGRA.
  Interestingly, we also find that tuning the $\betac$ parameter in
  the $\Rl$-$\Rh$ model can mimic some of the observational features
  of the \SPO model.
  This suggests a possible degeneracy between physical and parametric
  prescriptions.
\end{abstract}


\section{Introduction}
\label{sec:intro}

The Event Horizon Telescope~(EHT) enables the study of black hole
accretion flows on horizon scales through direct
imaging~\citepalias{2019ApJ...875L...1E, 2022ApJ...930L..12E}.
By using Very Long Baseline Interferometry~(VLBI), the EHT effectively
creates a virtual Earth-size telescope and achieves an angular
resolution of approximately 20\,$\mu$as.
Its images of the supermassive black holes at the centers of
Messier~87~(hereafter M87*) and the Milky Way~(hereafter \SGRA) have
opened new ways for testing general relativity in extreme gravity and
for probing plasma physics near event horizons.

The primary targets of EHT, M87* and \SGRA, with Eddington ratios
around $10^{-5}$~\citepalias{2019ApJ...875L...5E} and
$10^{-9}$~\citepalias{2022ApJ...930L..16E} respectively, are several
orders of magnitude less luminous than typical active galactic nuclei.
These radiatively inefficient accretion systems have very low density,
so the radiative cooling time scale is significantly longer than the
accretion time scale.
They are often modeled by, e.g., Advection Dominated Accretion
Flow~(ADAF), in which most of the dissipated energy is advected into
the black hole rather than radiated away~\citep{1994ApJ...428L..13N,
  1995ApJ...452..710N, 2014ARA&A..52..529Y}.

At such low densities, Coulomb collisions are too rare to thermalize
the electrons and ions, placing the plasma in the collisionless
regime~\citep[see, e.g.,][and references
  therein]{2023ApJ...959L...6T}.
Furthermore, electrons cool more efficiently than ions due to
synchrotron radiation.
This naturally leads to a two-temperature plasma, in which ions and
electrons maintain distinct temperatures.
A two-temperature ADAF was, in fact, the first model to successfully
explain the observed spectrum of \SGRA~\citep{1995Natur.374..623N}.
Since the observable emission arises from electrons, accurate modeling
of the electron distribution function and temperature is crucial for
interpreting EHT images~\citepalias{2022ApJ...930L..16E}.

In the last decade, general relativistic magnetohydrodynamics (GRMHD)
has become the workhorse for simulating black hole accretion systems
in all regimes~\citep{2000ApJ...528..462H, 2003ApJ...589..444G}.
The standard algorithm treats the accretion plasma as a fluid with a
single temperature.
To model the two-temperature plasma around M87* and \SGRA, it is
common to assume an ion-electron temperatures ratio $R \equiv T_i /
T_e$ in post-processing to map the electron temperature from the
plasma internal energy.
Then, using general relativistic ray tracing (or radiative transfer,
GRRT), one can simulate images of black
holes~\citep{2009ApJ...703L.142D, 2009ApJ...706..497M,
  2013ApJ...777...13C}.

However, GRMHD simulations naturally produce regions with contrasting
magnetic properties, leaving little flexibility for adjustment.
It soon became clear that adopting a constant $R$ across the whole
computational domain is insufficient to generate synthetic images that
reproduce the observed data~\citep{2015ApJ...799....1C,
  2015ApJ...812..103C}.

Early models of Sgr~A* treated the electrons in the disk and jet
separately to study how the jet contributes to the radio
emission~\citep{2013A&A...559L...3M, 2014A&A...570A...7M}.
\citet{2015ApJ...799....1C, 2015ApJ...812..103C} used plasma $\beta
\equiv P_\mathrm{gas}/P_\mathrm{mag}$ to create models that depend
only on local plasma parameters to capture electron acceleration and
cooling.
In the low $\beta$ region where the magnetic field is strong, the
electron temperature can either be a constant, or the ion-electron
ratio can be a constant.

Currently, the most common prescription was introduced by
\citet{2016A&A...586A..38M}, in which the ion-to-electron temperature
ratio is given by
\begin{equation}
  R(\beta) = \Rh\frac{(\beta/\betac)^2}{1+(\beta/\betac)^2}+
  \Rl\frac{1}{1+(\beta/\betac)^2}.
  \label{eq:RhRl}
\end{equation}
Here, $\Rh$ denotes the highest attainable temperature ratio in high
$\beta$, low magnetic field, ``disk'' regions, while $\Rl$ denotes the
lowest attainable temperature ratio in low $\beta$, high magnetic
field, ``funnel'' regions.
The parameter \(\beta_{\rm crit}\) sets the location of the transition.
At $\beta=\beta_{\rm crit}$, $R=(R_{\rm low}+R_{\rm high})/2$.

This $\Rh$-$\Rl$ prescription is flexible enough to produce
simulations that can explain many of the observed phenomena in M87*
and \SGRA~(\citealt{2016A&A...586A..38M};
\citetalias{2019ApJ...875L...5E, 2022ApJ...930L..16E}).
Because this model relies on free parameters that are not directly
constrained by first-principles physics, there is no well-justified
way to choose specific values for a given system.
As a result, the EHT collaboration constructed large simulation
libraries that span a broad range of parameter combinations in order
to compare with observations \citepalias{2019ApJ...875L...5E,
  2022ApJ...930L..16E}.
This brute-force approach is computationally expensive and limits the
efficiency of model selection.
These challenges have motivated the development of physically
motivated electron temperature models that aim to predict the
ion-electron temperature ratio from underlying plasma
physics~\citep{2015MNRAS.454.1848R, 2017MNRAS.466..705S,
  2018MNRAS.478.5209C}.

Building upon previous work, \citet*{2023ApJ...955...47S,
  2024ApJ...969..100S} proposed a new physically motivated electron
heating model (hereafter \SPO).
To capture how turbulent heating affects the electron-ion temperature
profile, \SPO is a covariant semi-analytic model that incorporates the
effect of Coulomb collisions, compressive heating, and radiative
cooling.
They also explored how the ratio of the driving powers of slow
compressive waves and Alfv\'en waves $P_\mathrm{S}/P_\mathrm{A}$ and
the magneto-rotational instability \citep[MRI,][]{1991ApJ...376..214B,
  2006MNRAS.372..183P} are connected.
By integrating small-scale turbulence to solve for global electron
thermodynamics of accretion flows, the \SPO model is in some sense a
Reynolds-Averaged Navier-Stokes~\citep[RANS,][]{1895RSPTA.186..123R}
formulation for electron thermodynamics.

We investigate the observational effects of the \SPO model by
implementing it in the general relativistic radiative transfer~(GRRT)
code \ipole~\citep{2018MNRAS.475...43M}.
We also discuss the effect of this model on black hole images to build
a clearer understanding of how observables~(such as variability) are
affected by energy partitioning in the accretion disk.

This paper is organised as follows: in section~\ref{sec:formulation},
we describe the formulation of the \SPO model, justifying its use it
in GRRT of GRMHD simulations, and validate its assumptions in GRMHD
simulations.
In section~\ref{sec:tests}, we perform tests of our implementation and
show the effects of different radiative transfer parameters.
In section~\ref{sec:application}, we study the impact of the \SPO
model on black hole images and compare the images with EHT
observations.
Finally, in section~\ref{sec:discussions}, we summarize our findings
and comparisons involving the \SPO model.

\section{A Local Electron Temperature Model for Accretion}
\label{sec:formulation}

We aim to replace the parametric prescription in
equation~\eqref{eq:RhRl} with a physically motivated one, while
keeping the rest of the GRRT framework unchanged so the multiple
models are directly comparable.
We also retain the assumption that $R$ depends solely on plasma
$\beta$.

\subsection{Turbulent Heating and the Ion-to-Electron Temperature Ratio}
\label{subsec:formulation_eq}

From the volume-integrated energy conservation equation, we can
express the temperature ratio as
\begin{equation}
  \frac{T_i}{T_e}=\frac{\gamma_i^{\mathrm{eff}}-1}{\gamma_e^\mathrm{eff}-1}\;
  \frac{1-(\gamma_e^\mathrm{eff}-1)(2-n_R)}{1-(\gamma^\mathrm{eff}_i-1)(2-n_R)
  (1+\frac{2\zeta}{\beta})}\left(\frac{Q_i}{Q_e}\right),
  \label{eq:SPO_R}
\end{equation}
where $n_R$ and $q$ are the power-law indices of the radial and
angular velocities, respectively.
$Q_i/Q_e$ is the ratio of energy dissipation between ions and
electrons.
$\zeta$ is the fraction of thermal pressure contributed by turbulence.
The adiabatic indices of ions and electrons are $\gamma_i$ and
$\gamma_e$, respectively.
The effective ion adiabatic index in the presence of turbulence is
$\gamma_i^{\rm eff}$.
This equation replaces equation~\eqref{eq:RhRl} to incorporate
microphysical effects on energy partition.
We describe the microphysics underlying this expression below.
The next two subsections describe the velocity profiles and the
adiabatic indices, respectively.

The energy partition mechanisms vary according to plasma $\beta$.
In high-$\beta$ region, ions receive energy from turbulence by Landau
damping.
In magnetically dominated, low-$\beta$ regions, energy partition is
dominated by MRI-induced compressive waves and Alv\'enic waves.
This $\beta$-dependent relationship can be captured by
\begin{equation}
\label{eq:4}
  \frac{Q_i}{Q_e}=\frac{35}{1+(\beta/15)^{-1.4}}+
  \frac{P_\mathrm{comp}}{P_\mathrm{A}},
\end{equation}
where $P_\mathrm{comp}/P_\mathrm{A}$ is the ratio of the driving
powers of compressive and Alfv\'enic waves.


The two linear compressive modes in MHD are slow magnetosonic waves
and fast magnetosonic waves.
If the fluctuation frequencies $\omega$ in the fluid are of the same
order as the Alfv\'en-wave frequency, the fast wave is negligible as
its timescale is much shorter than $1/\omega$
(\cite{2022JPlPh..88c9011K}, and Appendix
in~\cite{2024ApJ...969..100S}).
Therefore, the driving power ratio in low plasma $\beta$ regions can
be approximated by the slow-to-Alfv\'en driving power ratio $\PSPA$.

As both compressive and Alfv\'enic waves are driven by the MRI,
$\PSPA$ can be inferred from MRI-driven angular momentum transport.
If the mean magnetic field in the accretion disk is mostly azimuthal
because of strong shear, the slow-wave displacements lie along the
direction of shear~(denote $y$ as the azimuthal direction).
As a result, the energy injected by slow compressive waves is also
mostly due to the shear.
With the same assumption, the Alfv\'en wave displacements lie
perpendicular to the direction of shear~(denote $x$ as the radial
direction).
Therefore, the power ratio $P_\mathrm{S}/P_\mathrm{A}$ can be
re-written as
\begin{equation}
  \left(\frac{P_\mathrm{S}}{P_\mathrm{A}}\right)_{\vec{B_0}\approx B_0\hat{y}}
  = \frac{1}{2} \left\{q\left[\frac{M_{xy}}{R_{xy}}+1\right]-2\right\},
  \label{eq:PSPA}
\end{equation}
where $M_{xy}$ and $R_{xy}$ are components of the Maxwell stress.

\cite{2006MNRAS.372..183P} found that the stress ratio can be
expressed using the wavenumber of the unstable mode $k_M$ and the
shearing rate $q$,
\begin{equation}
  \frac{M_{xy}}{R_{xy}} = 1 + \frac{2(2-q)}{k_M^2+\gamma(k_M^2)}.
\end{equation}
Using shearing box simulations, \cite{2024ApJ...969..100S}
approximated the dominant unstable mode by the fastest growing linear
mode, $k_{\max}$ with a factor of 0.42, i.e.,
\begin{equation}
  k_M = 0.42k_{\max}=0.42\frac{\sqrt{q(4-q)}}{2}.
\end{equation}
The corresponding growth rate of this linear mode is then given by
\begin{equation}
  \gamma(k_M)=\Bigl[ \sqrt{(2-q)^2+4k_M^2}-k_M^2+(2-q)\Bigr]^{1/2}.
\end{equation}

\subsection{Radial Inflow and Orbital Shear}
\label{subsec:acc_profile}

To model the plasma velocity profile, \citet{2023ApJ...955...47S,
  2024ApJ...969..100S} introduced two power-law coefficients $n_R$ and
$q$ to describe the velocity profile of the accreting plasma.
Here, $n_R$ is the power-law index of the radial velocity,
with $\eta$ as a free parameter
\begin{equation}
  v^r=\frac{u^r}{u^t}=-\eta\left(\frac{r}{r_\mathrm{ISCO}}\right)^{-n_R}.
\end{equation}
The dimensionless orbital shear parameter $q$ is defined as the
power-law index of the angular velocity $\Omega(r)=u^\phi/u^t$.
\begin{equation}
  q=-\frac{d\ln{\Omega(r)}}{d\ln{r}}.
\end{equation}

Secondary turbulent effects in GRMHD introduce fluctuations in the
fluid velocities, which hinder the general fluid motion.
Following \cite{1895RSPTA.186..123R}, theorem 2d states that to apply
the dissipation function to an unsteady flow, the accuracy increases
with the averaging timescale in obtaining the mean motions.
Despite the local grid cell implementation of
section~\ref{subsec:formulation_eq}, we use velocities averaged over
time (following theorem 2d) and azimuth (assuming spherical symmetry)
in calculating $n_R$ and $q$.

The values of $n_R$ and $q$ can be fitted using GRMHD simulations.
As the derivation in equation~\eqref{eq:PSPA} relies on the mid-plane
assumption, $n_R$ and $q$ are fitted to mid-plane values for
consistency.
We use part of the EHT GRMHD simulation
library~\citep{2022ApJ...930L..16E} created by the KHARMA code
\citep{2021JOSS....6.3336P, 2022ascl.soft10013P}.
Due to the scope of the shearing box simulation in
\cite{2023ApJ...955...47S, 2024ApJ...969..100S}, we only consider the
Standard and Normal Evolution~(SANE) simulations
\citep{2012MNRAS.426.3241N, 2013MNRAS.436.3856S}, which are
characterized by turbulence and comparatively low magnetic flux $\phi
\approx 1$.

\begin{figure*}[t]
  \centering
  \includegraphics[width=\textwidth]{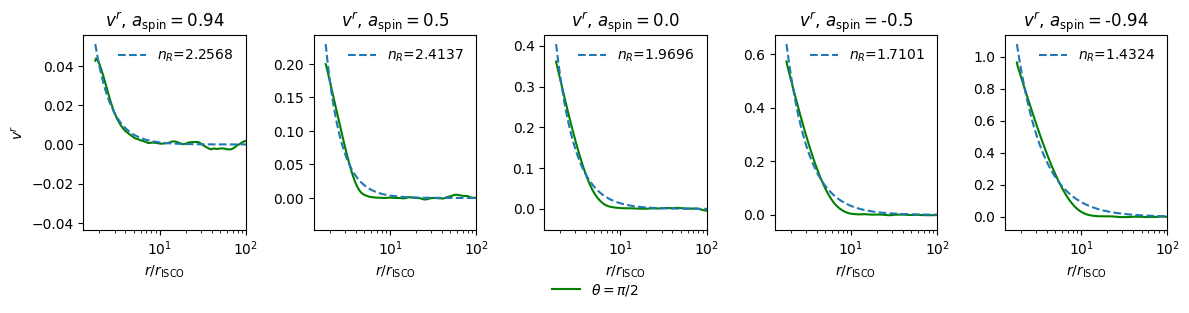}
  \includegraphics[width=\textwidth]{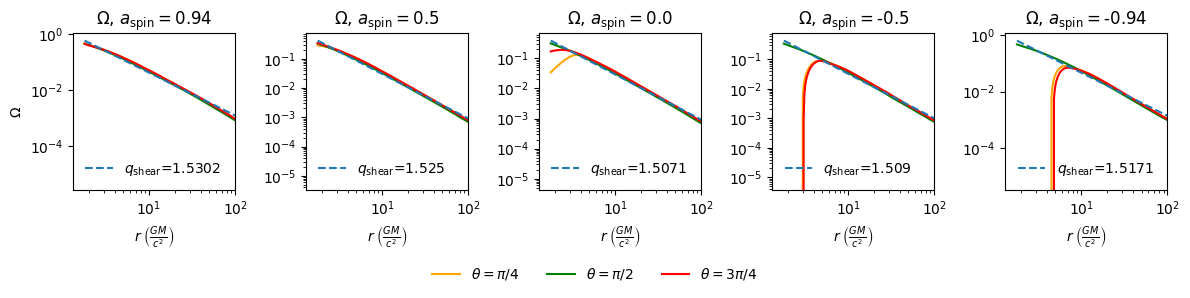}
  \caption{Temporal and azimuthal averages of $n_R$ and
    $q_\mathrm{shear}$, fitted to data from GRMHD simulations.
    To ensure that only values that have reached equilibrium are
    included, the analysis focuses on the inner accretion flow.
    For the radial velocity profile, only mid-plane values~($\theta =
    \pi/2$) are presented.
    For the angular velocity profile, values at $\theta = \pi/4$,
    $\pi/2$, and $3\pi/4$ are considered, as they achieve equilibrium
    values beyond the innermost stable circular orbit~($r_\mathrm{ISCO}$).}
  \label{fig:nR_qshear_plot}
\end{figure*}

Each GRMHD simulation uses $288\times128\times128$ grid cells in the
$(r,\theta,\phi)$ directions in modified spherical Kerr--Schild
coordinates, with $r$ extending to $1000\,GM/c^2$.
Five dimensionless spin values $a\equiv Jc/GM^2$ were used: $+0.94$,
$+0.5$, 0, $-0.5$, $-0.94$.
The sign of the spin denotes the alignment or anti-alignment of the
black hole spin and the angular momentum of the accretion flow.
To avoid the effect of turbulent fluctuations on $\Omega$ and $v^r$,
we compute the temporal and azimuthal averages of these two parameters
from 25,000--30,000\,$GM/c^3$.
As the outer accretion flow has not reached equilibrium, we only
consider the fitted values in the inner accretion flow.
The values are presented in Table~\ref{tab:nR_qshear_tab} and
Figure~\ref{fig:nR_qshear_plot}.

\begin{table}[t]
  \centering
  \begin{tabular}{c | c c c c c}
    \toprule
    \textbf{Spin} & \bm{$+0.94$} & \bm{$+0.50$} & \bm{$0.00$} & \bm{$-0.50$} & \bm{$-0.94$} \\
    \midrule
    $n_R$ & 2.2568 & 2.4137 & 1.9696 & 1.7101 & 1.4324 \\
    $q$   & 1.5302 & 1.5250 & 1.5071 & 1.5090 & 1.5171 \\
    \bottomrule
  \end{tabular}
  \caption{Fitted mid-plane values of $n_R$ and $q$ for GRMHD simulations with
    different spin parameters.}
  \label{tab:nR_qshear_tab}
\end{table}

It is clear that as the spin values decreases, $n_R$ parameter
approaches 1.5, which gives a Newtonian
profile~\citep{2022ApJ...941...88O}.
The shearing rates at the mid-plane show a near-Keplerian profile,
which is also consistent with the
literature~\citep[e.g.,][]{2021ApJ...914...55W, 2025ApJS..277...16D}.


\subsection{Adiabatic Indices and Electron Temperature}
\label{sec:Te_eff}

The adiabatic indices enter the equation of state of the fluid.
The \texttt{ipole} settings use $\gamma_i=5/3$ for ions and
$\gamma_e=4/3$ for electrons\footnote{We are aware that the fluid
adiabatic index is $\Gamma=4/3$ in KHARMA.
However, with Eq. (29) in \cite{2025ApJ...980..193G}, $\Gamma=4/3$ and
$\gamma_i=5/3$ yield $\gamma_e=11/9$, which is physically impossible,
and vice versa.
Therefore, we keep the default values of \ipole as a stopgap.}.
However, these indices are further modeled to capture the local
sub-grid effects.

Ions in the plasma are subject to turbulent effects, which we assume
scale linearly with the local Alv\'enic velocity.
This effect can be expressed by the effective ion adiabatic index
\begin{equation}
   \frac{1}{\gamma_i^\mathrm{eff}-1}=\frac{1}{\gamma_i-1}
   \left(1+\frac{2\zeta}{\beta}\right).
\end{equation}


Although the effect of turbulence on electrons is negligible because
$m_i>>m_e$, $\gamma_e$ is closely related to the speed of electrons.
In the nonrelativistic regime, the adiabatic index of electrons is
$\gamma_e=5/3$.
However, in the inner accretion flow, the electron temperatures become
relativistic due to the strong heating.
The electron adiabatic index decreases from $5/3$ to $4/3$ in regions
of high $\beta$.
To address this behavior, the empirical relationship for $\gamma_e$ is
modeled as follows, where $\beta_\mathrm{break}=5$ and $\kappa=0.13$.
\begin{equation}
  \gamma_e^\mathrm{eff} = \frac{4}{3} + \kappa\frac{(\beta/\beta_{\mathrm{break}})^2}
  {1+(\beta/\beta_\mathrm{break})^2}.
\end{equation}

\begin{figure}
  \centering
  \includegraphics[width=\columnwidth]{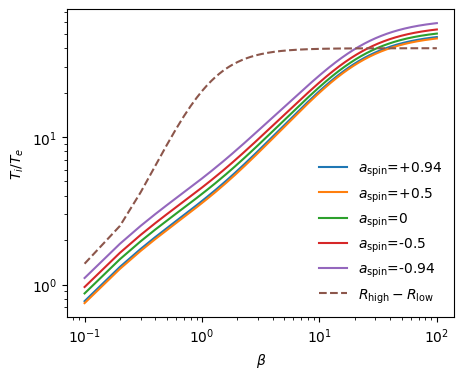}
  \caption{The ion-electron temperature ratio as a function of plasma
    $\beta$ for the standard $\Rh$-$\Rl$ prescription~(brown dashed
    line) and the \SPO model for $a_\mathrm{spin}=\pm0.94$, $\pm0.5$, 0.}
  \label{fig:temp_ratio}
\end{figure}

The \SPO temperature ratio profile is plotted in
Figure~\ref{fig:temp_ratio}.
At high $\beta$, the \SPO model reaches a plateau at
$T_i/T_e\approx40$, which shows good agreement with the EHT best-fit
SANE model.
On the other hand, \SPO demonstrates $T_i/T_e<1$ behaviour in the
low-$\beta$ region.
This is a result of the turbulent cascade, which is yet to be
constrained by observation.

In the \ipole calculation, the dimensionless electron temperature
$\Theta_e \equiv kT_e/m_ec^2$ is given by
\begin{equation}
  \Theta_e = \frac{m_i}{m_e}
  \frac{(\gamma_e-1)(\gamma_i-1)}{(\gamma_i-1)+(\gamma_e-1)\frac{T_i}{T_e}}
  \frac{U}{\rho}.
  \label{eq:Theta_e}
\end{equation}
Here, we use $\gamma_i$ and $\gamma_e$ instead of
$\gamma_i^\mathrm{eff}$ and $\gamma_e^\mathrm{eff}$ in
Eq~\eqref{eq:Theta_e}.
While $\gamma_i^\mathrm{eff}$ and $\gamma_e^\mathrm{eff}$ capture
sub-grid features within each GRMHD grid cell, $\Theta_e$ is computed
by global conservation of energy and density.
$\gamma_i$ and $\gamma_e$ provide a more suitable description for the
global plasma properties.

\begin{figure*}
  \centering
  \includegraphics[width=\textwidth]{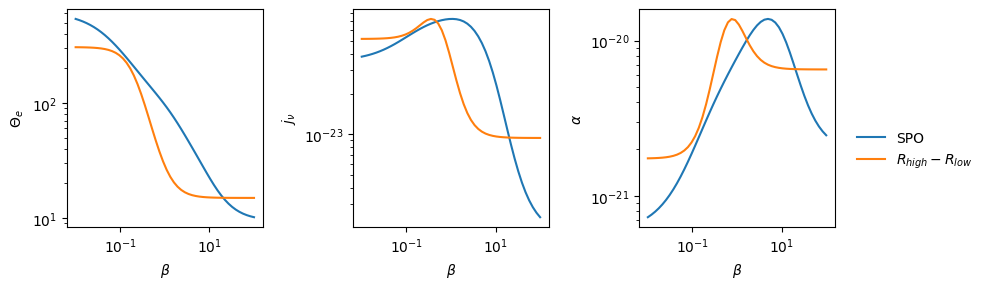}
  \caption{The dimensionless electron temperature $\Theta_e$,
    emissivity $j_\nu$, and absorptivity $\alpha$ for the \SPO
    model~(blue curve) and the $\Rl$-$\Rh$ model~(orange curve) are
    shown in the figure.
    The curves demonstrate that the two models differ both in the
    transition region $\beta\sim1$ and in their asymptotic behavior.}
  \label{fig:one-zone}
\end{figure*}

\begin{figure*}
    \centering
    \includegraphics[width=1\textwidth]{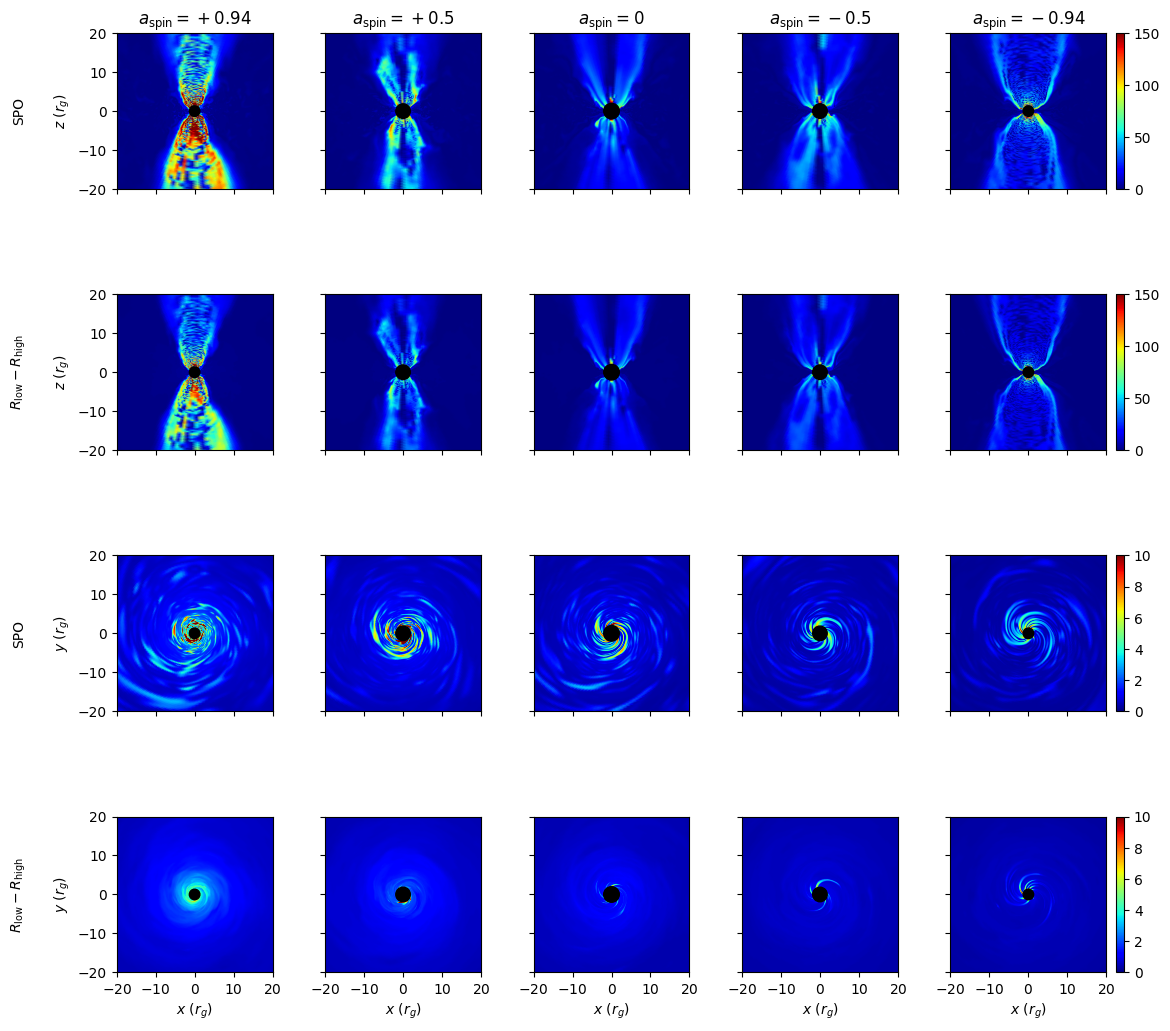}
    \caption{The electron temperature is computed for the SANE
      snapshot at $t=5000\,GM/c^3$ for $a=0.94$, 0.5, 0, $-0.5$, and 0.94.
      The first two rows show the $xz$ plane for the \SPO and
      $\Rl$-$\Rh$ models, respectively.
      The last two rows show the $xy$ plane for the same models in the
      same order.}
    \label{fig:Te}
\end{figure*}

\section{Implementation and Verification}
\label{sec:tests}

\subsection{\ipole Implementation}


We implemented the equations from section~\ref{sec:formulation} in the
electron model of \ipole.
This implementation is compatible with the current \ipole algorithm,
including the emissivity, absorptivity, and rotation coefficient along
the geodesic in the trace file.
Users can modify the runtime parameters $n_R$ and $q$ according to the
corresponding GRMHD simulation.
This implementation is available publicly
\footnote{\href{https://github.com/yttcari/ipole}
{https://github.com/yttcari/ipole}}.

\subsection{Convergence with Image Resolution}

\begin{figure*}
  \centering
  \includegraphics[width=0.8\textwidth]{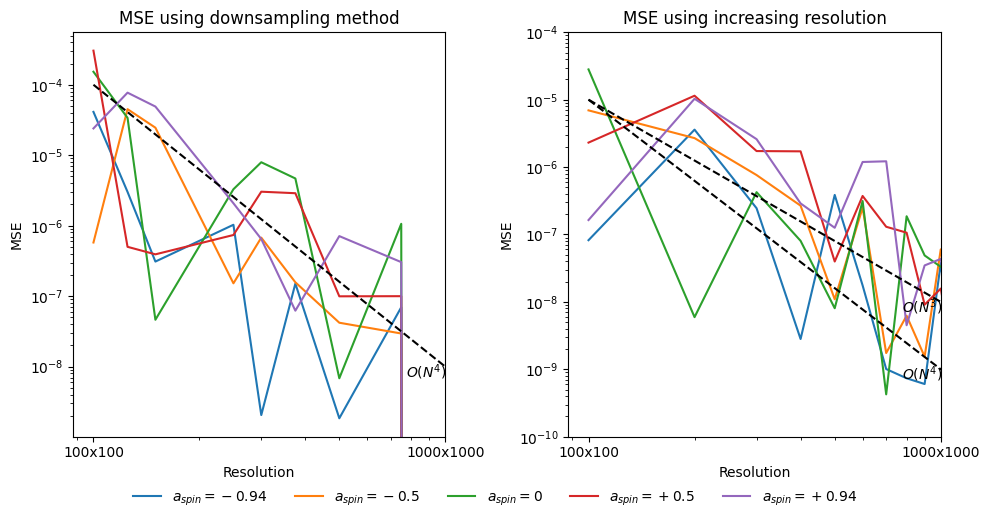}
  \caption{Self-convergence test: mean square error of \SPO images at
    different resolutions compared to high-resolution reference images.
    The difference curves, from blue to purple, represent
    $a_\mathrm{spin} = 0.94$, 0.5, 0, $-0.5$, $-0.94$, respectively,
    which follows the color convention of Figure~\ref{fig:temp_ratio}.
    The black dashed line shows $\mathrm{MSE} \propto N^{4}$, which
    shows second-order convergence for two-dimensional images.
  }
  \label{fig:convergence}
\end{figure*}

We do not expect a new electron temperature model to break \ipole.
However, because the dominant emitting region changes, the convergence
properties of the radiative transfer calculation may also change,
especially near the photon rings.
We employed two methods to evaluate the convergence properties of the
\SPO model in \ipole.
The first method is to generate an image at each of several
resolutions, ranging from $100\times100$ to $1000\times1000$ for the
five spin values.
For each resolution, we compute the mean squared error~(MSE) of the
total intensity relative to that of the reference image at
$1500\times1500$ resolution.
The second method is to first down-sample the $1500\times1500$
reference image, then compute the MSE of the total intensity relative
to the reference intensity.
We use this approach because the photon ring is due to the fact that
photon ring is a highly variable source.
When the resolution is not high enough to resolve the thin photon
ring, the resultant images may suffer from aliasing.
The MSEs for such a ``self-convergence test'' are plotted in
Figure~\ref{fig:convergence}.

\begin{figure*}
  \centering
  \includegraphics[width=1\linewidth]{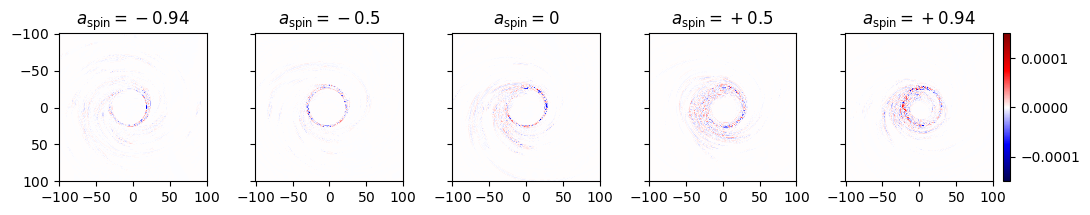}
  \caption{Intensity differences between the $200\times200$ and
    $1400\times1400$ images are shown in the figure.
    The $1400\times1400$ images are first down-sampled to
    $200\times200$ using blockwise averaging.}
  \label{fig:I_diff_resolution}
\end{figure*}

Both graphs show oscillations before converging toward the reference
image intensity across all five spin values.
From Figure~\ref{fig:I_diff_resolution}, it is clear that the
fluctuations originated from the photon rings.
This is expected, as the geodesics and hence the sampling of the fluid
are very sensitive to the impact parameter of the pixel.
The intensity variations are approximately $\pm$ $10^{-3}$ to
$10^{-4}$ Jy, with prograde spin black holes showing larger
fluctuations.
This is also expected, as the photon orbits in prograde black holes
are much closer to the event horizon than are the photon orbits in
retrograde black holes.

\subsection{Effect of the Magnetization Cutoff}

\begin{figure}
  \centering
  \includegraphics[width=1\linewidth]{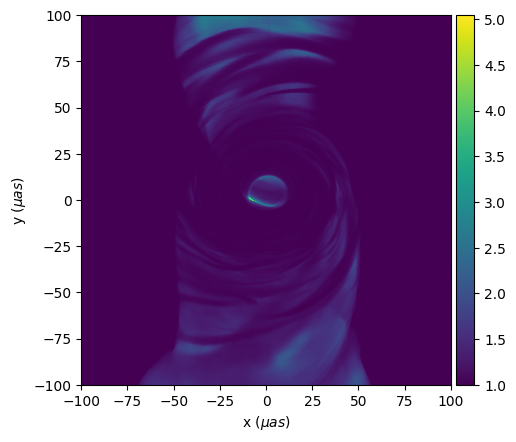}
  \caption{Intensity ratio between $\sigma_\text{cut}=50$ and
    $\sigma_\text{cut}=1$.
    The main difference in emission is located in the jet region, with
    a radially increasing trend.}
  \label{fig:sigma_cut}
\end{figure}

As the jet region is highly magnetically dominated, this region
suffers from numerical instability.
Errors accumulate and lead to nonphysical values of conserved
quantities.
To maintain a stable GRMHD simulation, an artificial ceiling is
introduced.
With the funnel magnetization defined as $\sigma \equiv B^2/\rho$,
assuming that emission from the jet region is negligible because of
its low density, \ipole sets an upper limit of $\sigma_\text{cut}=1$
by injecting artificial mass \citep{2019MNRAS.486.2873C,
  2021ApJ...910L..12E, 2024ApJ...964...17C}.

To investigate the effect of $\sigma_{\rm cut}$ in the \SPO model, we
generate images of a black hole with $a_{\rm spin}=0.94$ using
$\sigma_{\rm cut}=1$ and $\sigma_{\rm cut}=50$.
The intensity ratio of the two images is shown in
Figure~\ref{fig:sigma_cut}.
The total intensity of the image increased by 1.70\%, with most of the
increase in emission coming from the jet region and the black hole
shadow region.
However, because the jet emits little compared with the accretion
disk, the change is small relative to the overall emission.

\section{Application to \SGRA}
\label{sec:application}

Motivated by the reduced number of parameters in the \SPO model, we
now apply this new model in the context of \SGRA.
In \cite{2022ApJ...930L..16E}, there are no models that satisfy all
EHT and non-EHT constraints, with the tightest constraints being the
3-hour modulation index~($M_{\Delta T}$) and the 86\,GHz size
constraint.
Most of the models that fail only one constraint are MAD.
It would be interesting to demonstrate the effect of different
electron models on the constraints.
This would also help us better understand the physical meaning of the
parameters in the $\Rl$-$\Rh$ model.

\subsection{Effective $\Rl$-$\Rh$ Model}
\label{sec:RhRl_eff}

\begin{figure}
  \centering
  \includegraphics[width=1\columnwidth]{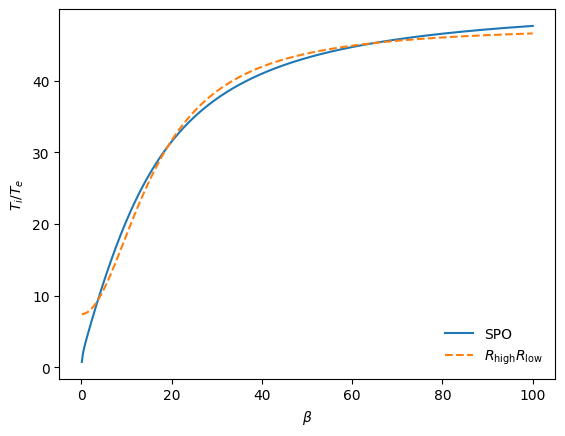}
  \caption{Effective $\Rh$-$\Rl$ temperature fitted based on the
    temperature computed using the \SPO model, with parameters
    $\Rh=47.60$, $\Rl= 7.43$, $\betac=16.22$.}
  \label{fig:effective_RhRl}
\end{figure}

Using the electron temperature shown in Section ~\ref{sec:Te_eff}, we
obtain the best-fit $\Rl$-$\Rh$ model, whose total flux is similar to
that of the \SPO model.
The best-fit parameter values for the $\Rh$-$\Rl$model, obtained from
the temperature computed using the \SPO model, are $\Rh=47.60$, $\Rl=
7.43$, $\betac=16.22$.
The best-fit value of $\Rh$ is similar to the values in most of the
best-fit SANE models in~\cite{2022ApJ...930L..16E}.
However, the values of $\Rl$ and $\betac$ are higher than the commonly
used values of $\Rl=1$ and $\betac=1$.

\subsection{Electron Temperature and Image Morphology}
\label{sec:image}

Even without computing the images, we can obtain some theoretical
expectations from the \SPO model.
Assuming constant values of $U=10$, $\rho=10$, we plot the electron
temperature, emissivity, and absorptivity following
Equation~\eqref{eq:Theta_e} and~\citet{2011ApJ...737...21L,
  2016ApJ...822...34P}.
The results are shown in Figure~\ref{fig:one-zone}.


The emission region is sensitive to the electron model.
From Figure~\ref{fig:one-zone}, we can see that the emitting regions
are closely related to the electron temperature.
The dominant emitting region in both models is at low $\beta$.
We expect more concentrated emission from the \SPO model because it
assigns a larger share of the energy to electrons.

Given the correlation between electron temperature and the emitting
region, we plot the electron temperature in the GRMHD simulations to
account for variations in $U$ and $\rho$.
Figure~\ref{fig:Te} shows the electron temperature from the GRMHD
simulation in the $xy$ and $xz$ planes.
Both models agree that the electron heating is more prominent in the
inner accretion flow and the jet, with a stronger effect in the \SPO
model.
The \SPO model is also more sensitive to fluctuation in $\beta$, which
reveal the spiral pattern in the inner accretion flow.
This phenomenon can be explained by the consideration of decreasing
resonance between Alfv\'en waves and ions, as discussed in
section~\ref{sec:formulation}.

\begin{figure*}
    \centering
    \includegraphics[width=1\linewidth]{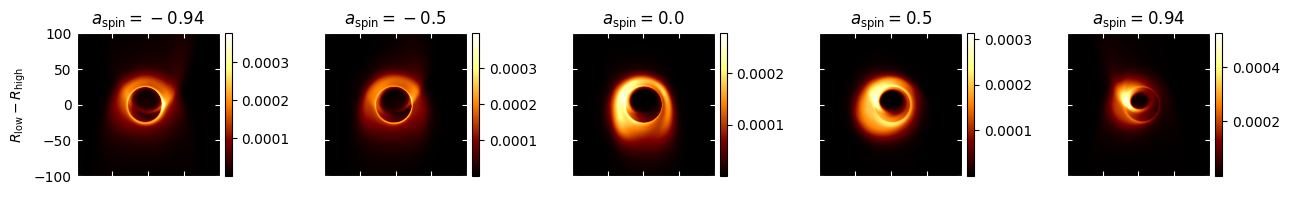}
    \includegraphics[width=1\linewidth]{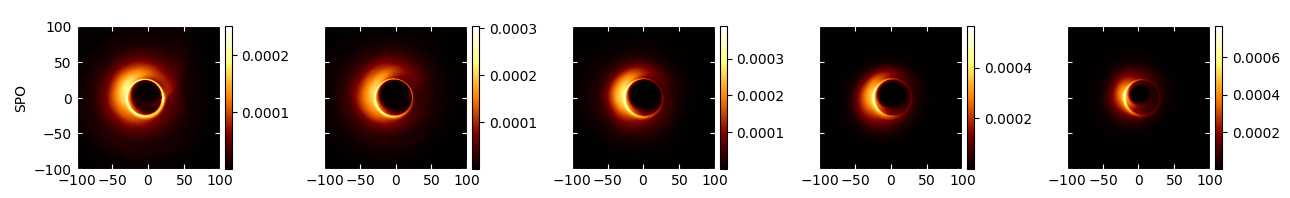}
    \caption{Time-averaged intensity images produced using \SPO model
      and SANE best-bet $\Rl$-$\Rh$ respectively.
      Each image is scaled separately.}
    \label{fig:images}
\end{figure*}

Figure~\ref{fig:images} shows the temporally averaged SANE images of
the \SPO model in the time range of 25,000--30,000\,$GM/c^3$.
The images show a clear circular black hole shadow.
The optical thickness shows a clear dependence on black hole spin.
The \SPO model produces images with lower optical thickness than the
$\Rl$-$\Rh$ model, especially for prograde spins.
To quantify the structural differences, we utilize the \texttt{REx}
function in \texttt{ehtim} to compute the $m$-ring diameter $d$,
$m$-ring width $w$, and the asymmetry $A$ of the time-averaged image.
The results for the two models are shown in Table~\ref{tab:ringfit}.

\SPO model produces a more asymmetric emission than $\Rl$-$\Rh$ model.
We observe that the asymmetric nature is closely related to the high
electron temperature region as denoted in Figure~\ref{fig:Te}.
The \SPO model partitions more energy to electrons in low-$\beta$
regions, making the inner accretion disk the primary emitting region.
Emission from the inner accretion disk alone is sufficient to produce
the targeted 2.4 Jy flux.
A similar relationship can also be seen in the $\Rl$-$\Rh$ model when
$\spin=0.94$.
Conversely, as the emitting region of the $\Rl$-$\Rh$ model is
concentrated at intermediate $\beta$, it produces a clear funnel other
than the accretion disk near the black hole.
This also results in more symmetric images in the $\Rl$-$\Rh$ model.

\begin{table}[!b]
  \centering
  \begin{tabular*}{\columnwidth}{@{\extracolsep{\fill}}lcccc@{}}
    \hline
    \textbf{Model} & \textbf{Spin} & \bm{$d\ (\mu\mathbf{as})$} & \bm{$w\ (\mu\mathbf{as)}$} & \bm{$A$} \\
    \hline
    {\SPO} & $-0.94$ & 45.30 & 7.37 & 0.41 \\
              & $-0.5$ & 46.72 & 4.70 & 0.47 \\
              & $0$  & 47.74 & 5.00 & 0.59 \\
              & $+0.5$  & 50.73 & 11.29 & 0.53 \\
              & $+0.94$  & 47.30 & 10.41 & 0.44 \\
    \hline
    {$\Rl$-$\Rh$} & $-0.94$ & 46.68 & 10.86 & 0.28 \\
                    & $-0.5$ & 46.31 & 9.94 & 0.29 \\
                    & $0$  & 42.42 & 20.37 & 0.37 \\
                    & $+0.5$  & 50.30 & 21.43 & 0.20 \\
                    & $+0.94$  & 40.47 & 16.62 & 0.46 \\
    \hline
  \end{tabular*}
  \caption{Ring model fit parameters for the \SPO and $\Rl$-$\Rh$
    models.}
  \label{tab:ringfit}
\end{table}

\begin{table*}[htbp]
  \centering
  \begin{tabular}{l c c c c c c}
    \toprule
    \textbf{Observational Constraint} & \textbf{Range} & \bm{$\spin=-0.94$} &
    \bm{$\spin=-0.5$} & \bm{$\spin=0$} & \bm{$\spin=0.5$} &
    \bm{$\spin=0.94$} \\
    \midrule
    $m_\mathrm{avg}$ (\%) & $(24, 28)$ & $7.11$ & $4.48$ & $2.63$ & $4.13   $ & $14.6$ \\
     $m_\mathrm{net}$ (\%) & $(2.0, 7.3)$ & $0.94$ & $0.90$ & $0.69$ & $1.13$ & $3.06$ \\
    $v_\mathrm{avg}$ (\%)& $(0.0, 5.5)$ & $0.18$ & $0.03$ & $-0.78$ & $-0.36$ & $-0.12$ \\
    $v_\mathrm{net}$ (\%)& $(-0.7, 0.12)$ & $-0.1$ & $-0.94$ & $-1.91$ & $-0.60$  & $-0.47$ \\
    $|\beta_1|$ & $(0.10, 0.14)$ & $0.01$ & $0.011$ & $0.009$ & $0.016$ & $0.061$ \\
    $|\beta_2|$ & $(0.14, 0.24)$ & $0.014$ & $0.013$ & $0.009$ & $0.018$ & $0.131$ \\
    $\angle \beta_2$ (deg) (observed) & $(125, 159)$ & $-47.47$ & $-77.31$ & $32.45$ & $68.55$ & $39.15$ \\
    $\angle\beta_2$ (deg) (RM derotated) & $(168, -85)$ &  &  &  &  &  \\
    $|\beta_2|/|\beta_1|$ & $(1.1, 2.1)$ & $1.40$ & $1.18$ & $1.00$ & $1.13$ & $2.16$ \\
    \bottomrule
  \end{tabular}
  \caption{Comparison of the observational constraints and the \SPO
    results at five values of of $\spin$.}
  \label{tab:obs_constraints}
\end{table*}

\subsection{Spectral Indices}

\begin{figure*}
  \centering
  \includegraphics[width=1\linewidth]{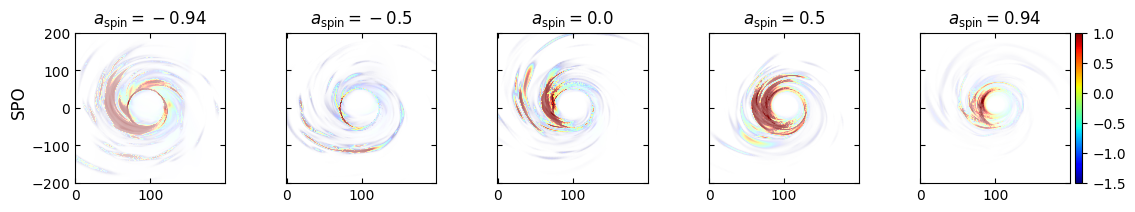}
  \includegraphics[width=1\linewidth]{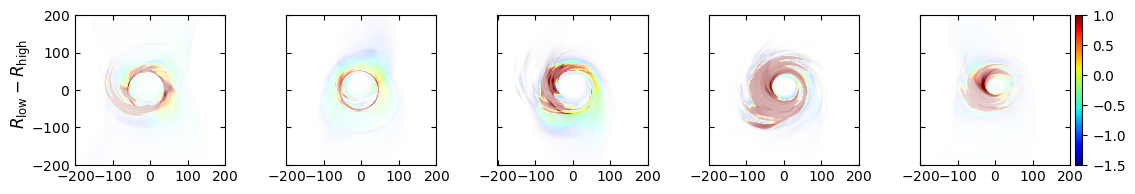}
  \caption{Spectral index maps for the \SPO and $\Rl$-$\Rh$ models.
    The opacity is scaled linearly according to the emission, with
    saturated opacity for the emission region with top 1\% intensity.}
  \label{fig:spec_index}
\end{figure*}

The spectral index $\alpha$ is related to several physical quantities
in the electron model, including magnetic field strength, optical
depth, and the electron energy
distribution~\citep{1986rpa..book.....R, 2023MNRAS.519.4203R}.
As the spectral index is sensitive to plasma heating mechanisms, it is
useful to investigate the spectral index of the \SPO model.
We computed the spectral map by
\begin{equation}
  \alpha=\frac{d\log{I}}{d\log{\nu}}
\end{equation}
Following the 2018 EHT observations, we use images at 214.1 and
228.1\,GHz to compute the spectral index map of the snapshot at
$t=25{,}000\,GM/c^3$, as shown in Figure~\ref{fig:spec_index}.
Both models show positive spectral indices around the photon ring.
In the $\Rl$-$\Rh$ model, these are concentrated in the forward funnel
region.
On the other hand, the negative spectral indices are concentrated in
the spiral-arm region.
These results are consistent with the analytical expectation
in~\cite{2023MNRAS.519.4203R}, where the spectral index peaks at the
photon ring and decreases radially in the spiral arms.
Furthermore, the positive spectral distribution matches the optically
thick regions, which is also intrinsically associated with the
electron temperature model.

Observations with the Atacama Large Millimeter/submillimeter Array~(ALMA)
give a spectral index of $0.0\pm0.1$~\citep{2023MNRAS.519.4203R,2022ApJ...930L..19W}.
We compute the intensity-weighted spectral index by
\begin{equation}
  \alpha=\frac{\int I(x,y) \alpha (x,y) dxdy}{\int I(x,y) dxdy}.
\end{equation}
We find that images from the \SPO model generally have more negative
spectral indices than those from the $\Rl$-$\Rh$ model, which can be
attributed to the extended spiral structure in the chosen snapshot.
However, no snapshot falls into the range of $\alpha=0\pm0.1$.
A more comprehensive study can be done by considering time-averaged
images.

\begin{table}[h]
  \centering
  \begin{tabular}{ccc}
    \hline
    \textbf{\boldmath $a_{\text{spin}}$} & \textbf{\boldmath \SPO} & \textbf{\boldmath $\Rl$-$\Rh$} \\
    \hline
    $-0.94$ & $0.11$  & $0.13$  \\
    $-0.50$ & $-0.87$ & $-0.30$ \\
    $0.00$  & $-0.04$ & $0.18$  \\
    $+0.50$ & $0.53$  & $0.98$  \\
    $+0.94$ & $-0.12$ & $0.35$  \\
    \hline
  \end{tabular}
  \caption{Spectral index values for the \SPO and $\Rl$-$\Rh$ models
    across different spin parameters ($a_{\text{spin}}$).}
  \label{tab:spectral_index_values}
\end{table}
\subsection{Three-Hour Flux Variability}

\begin{figure*}
  \centering
  \includegraphics[width=1\textwidth]{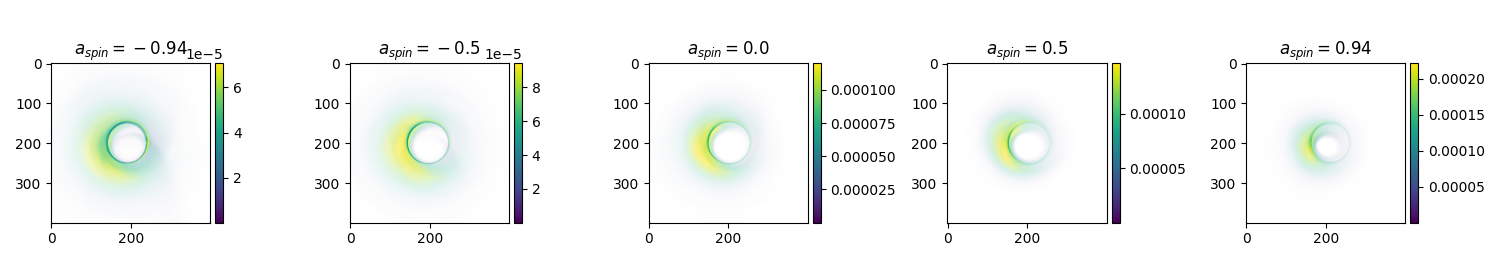}
  \caption{Maps of the standard deviation of the intensity over the
    simulation.
    For better visibility, the opacity is scaled according to the
    emission.}
  \label{fig:sdmap}
\end{figure*}

\begin{figure*}
  \centering \includegraphics[width=1\linewidth]{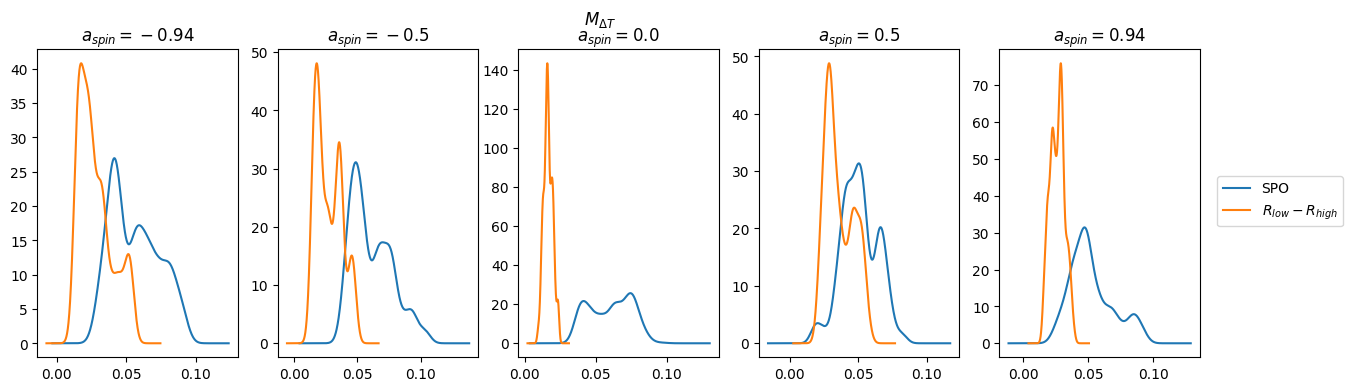}
  \caption{Kernel density estimates for the three-hour modulation
    index $M_3$.}
  \label{fig:M3}
\end{figure*}

The tightest constraint from the EHT observations is the 3-hour
modulation index $M_{\Delta T}\equiv \sigma_{\Delta T}/\mu_{\Delta
  T}$, where $\sigma_{\Delta T}$ and $\mu_{\Delta T}$ denote the
standard deviation and mean of the total flux over a three-hour
interval~(530 $GM/c^3$) respectively~\citep{2015ApJ...812..103C,
  2022ApJ...930L..16E}.
We consider a duration of 5000 $GM/c^3$, with equally spaced windows.
The resultant plot is shown in Figure~\ref{fig:M3}.

Currently, there is still a lack of studies on the physical effects of
modifying parameters $\Rl$ and $\betac$.
When $\Rl=1$, \citet{2024ApJ...964...17C, 2025ApJ...985..164C} show
that the main contribution to the variability comes from the photon
ring.
This effect is more prominent in prograde black holes, whereas
retrograde black holes demonstrate secondary effects due to the
misalignment between the accretion disk and the black hole spin.
It is possible to reduce the variability by varying the value of
$\Rl$.
We expect the effective $\Rl$-$\Rh$ model to produce a decrease in
$M_{\Delta T}$ for prograde black holes.

Generally, both the \SPO and $\Rl$-$\Rh$ models produce relatively
less variable emission, with the \SPO model slightly more variable
than the $\Rl$-$\Rh$ model~(Figure~\ref{fig:M3}).
This can be explained by the low optical depth of the \SPO model
\citep{2024ApJ...964...17C, 2025ApJ...985..164C}.
As the main source of variability comes from the photon ring, the
optically thick accretion disk created by the $\Rl$-$\Rh$ model blocks
the emission from the photon ring, resulting in less variable emission
overall.

\subsection{Polarization}

\cite{2024ApJ...964L..25E} present the first \SGRA polarization
analysis, which provides tighter constraints on the parameter space.
We summarize the observational constraints and our results in
Table~\ref{tab:obs_constraints}.
None of the \SPO models satisfy the full set of constraints for \SGRA.
This shows that both the electron temperature and the plasma
properties play a significant role in polarimetry.
Here, we will discuss some characteristic differences between the \SPO
and $\Rl$-$\Rh$ models in polarimetry.

The \SPO model generally produces lower linear and circular
polarization, while most of the emission detected comes from the jet
region that is closer to the inner accretion flow.
This result agrees with the intrinsically high Faraday depth of the
SANE model, where the accretion flow depolarizes most of the polarized
emission.

Comparing with $\Rl$-$\Rh$ model, $\Rl$-$\Rh$ model has a larger value
and a more concentrated value of $m_{avg}$ and $v_{avg}$ than \SPO
model, which can be attributed to the strong jet emission.
Although the jet is not clearly visible in Figure~\ref{fig:images},
the jet predicted by Section~\ref{sec:image} has a significant effect
on polarimetry.
The forward jet in the $\Rl$-$\Rh$ model behaves as a source of linear
polarization and positive circular polarization, resulting in a higher
polarization fraction despite the larger Faraday depth in the
$\Rl$-$\Rh$ model.

\section{Discussions}
\label{sec:discussions}

Electron heating mechanisms remain understudied in black hole physics.
\cite{2023ApJ...955...47S, 2024ApJ...969..100S} presented a
semi-analytical physically motivated electron temperature model.
Motivated by the reduction in the number of electron temperature model
parameters, we aim to study the effect of the \SPO model.
In this paper, we compared the images produced by the \SPO and
$\Rl$-$\Rh$ models in total intensity and polarization by performing
GRRT.
We then used \SPO model in the context of \SGRA to see whether the EHT
conclusion is sensitive to the electron heating model.

We obtain the global accretion parameters $n_R$ and $q$ by temporally
and azimuthally averaging the radial and orbital velocities in the
SANE GRMHD simulations.
While all of the SANE simulations give $q\approx1.5$, demonstrating a
near-Keplerian profile, there is a decreasing trend of $n_R$ according
to $|a_\mathrm{spin}|$.
However, we recognize that this is a simplified description of these
quantities.
From~\cite{2022ApJ...941...88O}, $v^r$ should be obtained by the
normalization condition $\mathbf{v}\cdot\mathbf{v}=-1$ at or inside
the ISCO radius.
We also only consider the fitted quantities inside the inner accretion
region, as the outer accretion flows have not reached equilibrium
because of the computational cost of the GRMHD simulations.

The \SPO model produces higher electron temperature in low-$\beta$
regions, such as the jet and inner accretion disk.
This is a direct consequence of the plasma turbulence cascade, which
is yet to be captured by the $\Rl$-$\Rh$ model.
For a fixed \SGRA intensity of 2.4 Jy, most of the emission in the
\SPO model comes from the inner accretion disk.
This results in an asymmetric, optically thin accretion disk and a
clear photon ring when compared to the $\Rl$-$\Rh$ model.

The appearance of the photon ring significantly affects the
variability and polarization of images produced by the \SPO model.
As the photon ring itself is a highly variable and polarized source,
an optically thin accretion disk reduces the extent of Faraday
derotation and depolarization, which increases the variability and
polarization of the \SPO images.
However, the resultant images still fall outside the observational
constraints.
This agrees with the EHT results that \SGRA is MAD, which illustrates
the innate role of plasma properties despite the influence of electron
models.

The \SPO model is formulated according to SANE simulations.
On the one hand, given that \SGRA and M87* are very likely MAD based
on EHT observations, it would be useful to expand the formulation to
MAD black holes.
On the other hand, the SANE formulation may be useful for describing
ion-electron behavior in radio-quiet black holes.
This aligns with some of the future targets of the
ngEHT~\citep{2025ApJ...985...41Z}, allowing verification of the \SPO
model.

Because of computational limitations, we consider only simulations
viewed at an inclination of $30^\circ$.
Overall, we agree with EHT conclusion that \SGRA is unlikely a SANE
black hole.
The observational data still favours MAD model.
Although plasma properties play a significant role in black hole
observables, electron models mainly affect the optical depth and
Faraday depth of the accretion disk, which are significant in
determining the image morphology, polarization properties and
variability of the black hole.
To provide a more general study of the effect of the \SPO model, it
would be interesting to see how face-on and edge-on views would affect
the intensity and polarization images.

The \SPO model provides a physically motivated description of plasma
heating in black hole accretion.
Compared with the $\Rl$-$\Rh$ model, it reduces the number of free
parameters, allowing more efficient use of computational resources.
This can provide us with further insight into black hole physics.

\begin{acknowledgments}
  We thank George Wong for helpful discussions and assistance with the
  GRMHD simulations and \texttt{ipole}.
  We used the EHT GRMHD simulation library, developed in part through
  C.K.C.'s Frontera allocation AST20023 at the Texas Advanced
  Computing Center.
  Frontera is supported by NSF award OAC-1818253.
  Y.T.T. acknowledges partial support from the CUHK SURE program.
  C.K.C. acknowledges partial support from NSF grant AST-2535855.
\end{acknowledgments}

\bibliographystyle{aasjournal}
\bibliography{refs,main}

\end{document}